\documentclass[twocolumn,aps,prl]{revtex4}
\usepackage{amsfonts}
\usepackage{amsmath}
\usepackage{amssymb}
\usepackage{graphicx}
\usepackage{color}
\usepackage{ulem}
\usepackage[colorlinks, urlcolor=cyan, citecolor=blue, linkcolor=magenta]{hyperref}

\begin{document}

\title{Gauge Violation Spectroscopy in Synthetic Gauge Theories}
\author{Hao-Yue Qi}
\author{Wei Zheng}
\email{zw8796@ustc.edu.cn}
\affiliation{Hefei National Laboratory for Physical Sciences at the Microscale and
Department of Modern Physics, University of Science and Technology of China,
Hefei 230026, China}
\affiliation{CAS Center for Excellence in Quantum Information and Quantum Physics,
University of Science and Technology of China, Hefei 230026, China}
\affiliation{Hefei National Laboratory, University of Science and Technology of China,
Hefei 230088, China}
\date{\today }

\begin{abstract}
Recently synthetic gauge fields have been implemented on quantum simulators.
Unlike the gauge fields in the real world, in synthetic gauge fields, the
gauge charge can fluctuate and gauge invariance can be violated, which
leading rich physics unexplored before. In this work, we propose the gauge
violation spectroscopy as a useful experimentally accessible measurement in
the synthetic gauge theories. We show that the gauge violation spectroscopy
exhibits no dispersion. Using
three models as examples, two of them can be exactly solved by bosonization,
and one has been realized in experiment, we further demonstrate the gauge
violation spectroscopy can be used to detect the confinement and
deconfinement phases. In the confinement phase, it shows a delta function
behavior, while in the deconfinement phase, it has a finite width.
\end{abstract}

\date{\today}
\maketitle

Gauge theories play a central role in modern physics. On one hand, gauge
theories provide a unified description of fundamental interactions between
elementary particles within the Standard Model~\cite{Weinberg@1995.BK}. On
another hand, gauge fields emerge from the low-energy effective theories of
strongly correlated condensed matter~\cite{Kogut@1979.RMP,Fradkin@1995.BK}. For example, the Chern-Simons gauge field can effectively
describe the behavior of fractional quantum Hall fluids \cite{Fradkin@1991.CS}. Gauge fields arise naturally as the slave-particle technique is
applied to the quantum magnets. In quantum information theory, Kitaev's
toric code is a $Z_{2}$ lattice gauge model~\cite{Kitaev@2003.TC}. Despite the
success of gauge theories, studying the real-time dynamics of gauge fields
is a notable challenge due to the limit of the classical computational
methods. To overcome these limitations, synthetic
gauge fields have been implemented on quantum simulators based on ultracold
atoms in optical lattices~\cite{Bloch@2019.LGT,YZS@2020.QLM,Hauke@2020.LGT,Yuan@2022.ETH,Yuan@2022.ETH&critical,Yuan@2023.angle}, trapped ions ~\cite{Zoller@2016.Ions}, or superconducting qubits
~\cite{Klco@2018.CS,Klco@2020.CS}.

\begin{figure}[tbp]
\centering
\includegraphics[width=0.45\textwidth]{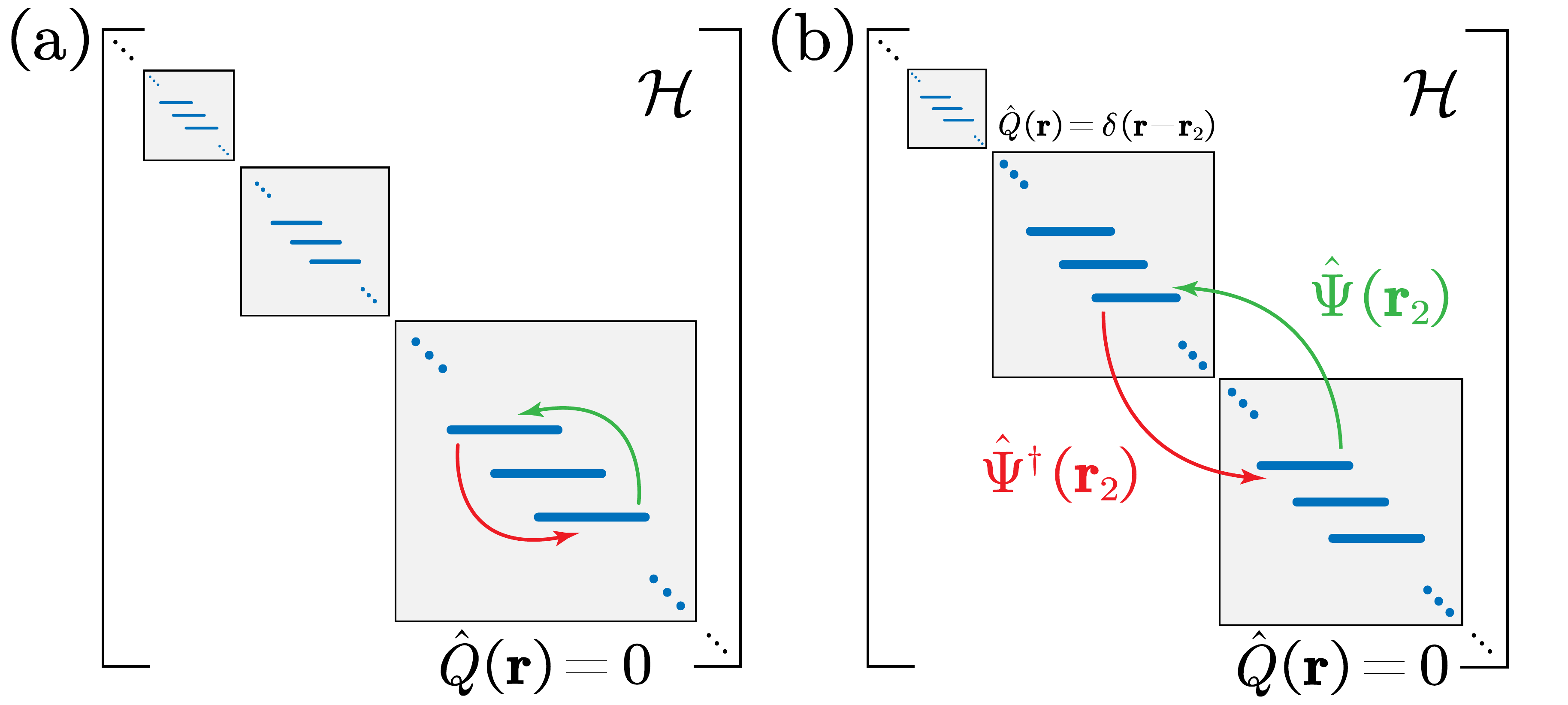}
\caption{Schematic of gauge invariance (a) and violation (b) spectroscopy.
The blocks represent different gauge sectors labeled by $\hat{Q}(\mathbf{r})$
in Hilbert space. In the gauge violation case (b), the operations $\hat{\Psi}%
(\mathbf{r}_2),\hat{\Psi}^\dag(\mathbf{r}_2)$ induce a transition between $%
\hat{Q}=0$ and $\hat{Q}=\protect\delta(\mathbf{r}-\mathbf{r}_2)$ gauge
sectors.}
\label{f1Hilbert}
\end{figure}

The key concept of gauge theories is the local gauge symmetry, $\left[ \hat{G%
}(\mathbf{r}),\hat{H}\right] =0$, where $\hat{G}(\mathbf{r})$ is the local
gauge transformation and $\hat{H}$ is the Hamiltonian of the system. Local
gauge symmetry separates the Hilbert space into disconnected sectors
labelled by local gauge charge $\hat{Q}(\mathbf{r})$, the generator of $\hat{%
G}(\mathbf{r})$, see Fig.\ref{f1Hilbert}. In real world, we are living in
the so-called physical sector with vanishing gauge charge $\hat{Q}(\mathbf{r}%
)=0$. Projecting into the physical sector enforces an extensive number of local
constraints between matter and gauge fields, which is nothing but the
Gaussian law. However, in synthetic gauge theories on quantum simulators,
local gauge charge $\hat{Q}(\mathbf{r})$ is not restricted to the physical
sector. It can even fluctuate due to inter-sector superposed initial states
or gauge violation perturbations~\cite{Jad@2020.GV,Jad@2020.noise,Jad@2020.reliability}. Such fluctuations lead to richer gauge
violation physics in synthetic gauge theories. For example, disorder-free
localization can emerge in synthetic gauge theories by preparing the initial
states as the superposition of several sectors~\cite{Smith@2017.Localization,Scardicchio@2017.Localization,Smith@2018.Localization,Jad@2023.Localization}. If the
Gauss's law is not imposed, the ground state of the $Z_{2}$ lattice gauge
theory forms the charge density wave in the non-physical sector \cite{Zheng@2020.LGT}.
Besides, allowing transitions between different sectors can also
lead to exotic phase transition that doesn't exist in real gauge theories
\cite{Assaad@2016.LGT}.

Spectroscopy measurement is a powerful technique to detect the excitations
in both real materials and quantum simulators. It has been widely used to
probe the spectrums of single-particle and collective excitations, which
determine the phases and dynamics of a given quantum system. For example,
the angle-resolved photoemission spectroscopy (ARPES)~\cite{Shen@2021.ARPES} has been applied to
study the pseudo gap of high-$T_{c}$ cuprates and edge states of topological
insulators. In ultracold atomic gases, radio frequency (RF) spectroscopy~\cite{CQJ@2009.RF} is
used to measure single-particle spectrums of Fermi gases~\cite{Jin@2008.RF,Ketterle@2008.RF,ZhangJin@2012.RF,Zwierlein@2012.RF,Zwierlein@2019.RF} and Fermi/Bose polarons~\cite{Zwierlein@2019.F-Polaron,Grimm@2012.F-Polaron,Roati@2017.F-Polaron,Bruun@2016.B-Polaron,Jin@2016.B-Polaron}. Several spectroscopy techniques have also been developed in other
quantum simulation platforms, such as trapped ions~\cite{Monroe@2014.Ions-Spectra,Roos@2015.Ions-Spectra} and superconducting
qubits~\cite{Angelakis@2017.SC-spectra}. However, the spectroscopy study of synthetic gauge theories on
quantum simulators is still missing.

In this paper, we propose the gauge violation spectroscopy in synthetic
gauge theories on quantum simulators. By gauge violation, we mean that the
measurement induces a transition between different gauge sectors, see Fig.%
\ref{f1Hilbert}. We demonstrate that the usual single-particle spectroscopy
measurement process (such as RF spectroscopy in ultracold atomic gases) of
synthetic gauge theories is gauge violation rather than gauge invariant. Besides, the gauge invariant spectroscopy needs highly non-local probes, and is challenging in current simulators. Furthermore, we show that the gauge violation spectroscopy
exhibits no dispersion, as it violates the Gauss law. Using three models that possess local $U(1)$ gauge symmetries as examples, we
show that the gauge violation spectroscopy can be used to detect the
confinement and deconfinement phases in gauge theories.
In the confinement phase, gauge violation spectrum is nearly a delta
function, while in the deconfinement phase, it exhibits a finite width.

\textit{Concept}. We know that the single-particle absorbing and emission
spectrum function can be obtained from the Fourier transformation of the
following Green's functions
\begin{eqnarray}
\mathcal{A}_{\mathrm{ab}}(\mathbf{k},\omega ) &=&i\int dt\int d\mathbf{r}g_{%
\mathbf{r}}^{>}(t)e^{-i(\mathbf{k\cdot r}-\omega t)}, \\
\mathcal{A}_{\mathrm{em}}(\mathbf{k},\omega ) &=&i\varepsilon \int dt\int d%
\mathbf{r}g_{\mathbf{r}}^{<}(t)e^{-i(\mathbf{k\cdot r}-\omega t)}.
\end{eqnarray}%
These Green's functions are defined as%
\begin{eqnarray}
ig_{\mathbf{r}_{2}-\mathbf{r}_{1}}^{>}(t) &=&\left\langle \psi
_{0}\right\vert \hat{\Psi}(\mathbf{r}_{2},t)\hat{\Psi}^{\dag }(\mathbf{r}%
_{1},0)\left\vert \psi _{0}\right\rangle , \\
ig_{\mathbf{r}_{2}-\mathbf{r}_{1}}^{<}(t) &=&\varepsilon \left\langle \psi
_{0}\right\vert \hat{\Psi}^{\dag }(\mathbf{r}_{1},0)\hat{\Psi}(\mathbf{r}%
_{2},t)\left\vert \psi _{0}\right\rangle ,
\end{eqnarray}%
where $\hat{\Psi}(\mathbf{r},t)=e^{iHt}\hat{\Psi}(\mathbf{r)}e^{-iHt}$ is
the matter field operator in Heisenberg picture, and $\varepsilon =+1(-1)$
for Bosonic (Fermionic) matter field. Here we choose $\left\vert \psi
_{0}\right\rangle $ to be the ground state in the physical sector, i.e. the
sector with vanishing gauge charge $\hat{Q}(\mathbf{r})=0$. These Green
functions describe the process that adding one particle (hole) to the system
at position $\mathbf{r}_{1}$, and removing one particle (hole) at position $%
\mathbf{r}_{2}$ after evolution time $t$. In synthetic gauge theories, the
gauge field can not adjust to follow the charge we add (remove). Thus this
process violates the Gaussian law, and excites the system away from the
physical sector. More specifically, the matter field operator is not invariant
under a local gauge transformation, thus $\left[ \hat{\Psi}\left( \mathbf{r}%
^{\prime }\right) ,\hat{Q}(\mathbf{r})\right] \neq 0$. One finds $\hat{Q}(%
\mathbf{r})\hat{\Psi}^{\left( \dag \right) }\left( \mathbf{r}^{\prime
}\right) \left\vert \psi _{0}\right\rangle =\pm e\delta (\mathbf{r}-\mathbf{r%
}^{\prime })\hat{\Psi}^{\left( \dag \right) }\left( \mathbf{r}^{\prime
}\right) \left\vert \psi _{0}\right\rangle $. Note that the state $\hat{\Psi}%
^{\left( \dag \right) }\left( \mathbf{r}^{\prime }\right) \left\vert \psi
_{0}\right\rangle $ is no longer in the physical sector as shown in Fig.\ref%
{f1Hilbert}(b). Then the Green's functions can be simplified into

\begin{eqnarray}
ig_{\mathbf{r}_{2}-\mathbf{r}_{1}}^{>}(t) &=&\delta (\mathbf{r}_{2}-\mathbf{r%
}_{1}\mathbf{)}\left\langle \psi _{0}\right\vert \hat{\Psi}(\mathbf{r}_{2}%
\mathbf{)}e^{-i\hat{H}_{-}t}\hat{\Psi}^{\dag }(\mathbf{r}_{1})\left\vert
\psi _{0}\right\rangle ,  \label{GV.CR01} \\
ig_{\mathbf{r}_{2}-\mathbf{r}_{1}}^{<}(t) &=&\varepsilon \delta (\mathbf{r}%
_{2}-\mathbf{r}_{1}\mathbf{)}\left\langle \psi _{0}\right\vert \hat{\Psi}%
^{\dag }(\mathbf{r}_{1})e^{+i\hat{H}_{+}t}\hat{\Psi}(\mathbf{r}_{2}\mathbf{)}%
\left\vert \psi _{0}\right\rangle .  \label{GV.CR02}
\end{eqnarray}%
where $\hat{H}_{\xi }\equiv \hat{H}\left[ \hat{Q}(\mathbf{r})=\xi e\delta (%
\mathbf{r-r}_{1})\right] $, $\xi =\pm $, is the Hamiltonian in the
non-physical sectors, and energy of $\left\vert \psi _{0}\right\rangle $ is
set to be zero. Here the delta-function $\delta (\mathbf{r}_{2}-\mathbf{r}%
_{1}\mathbf{)}$ is due to the fact that if $\mathbf{r}_{2}\neq \mathbf{r}_{1}
$, the state will not go back to the physical sector, $\hat{Q}(\mathbf{x})%
\hat{\Psi}\left( \mathbf{r}_{2}\right) \hat{\Psi}^{\dag }\left( \mathbf{r}%
_{1}\right) \left\vert \psi _{0}\right\rangle =\left\{ \delta \left( \mathbf{%
x-r}_{2}\right) -\delta \left( \mathbf{x-r}_{1}\right) \right\} \hat{\Psi}%
\left( \mathbf{r}_{2}\right) \hat{\Psi}^{\dag }\left( \mathbf{r}_{1}\right)
\left\vert \psi _{0}\right\rangle $. As a result, the gauge violation
spectrum exhibits no dispersion $\mathcal{A}_{\mathrm{ab/em}}(\mathbf{k}%
,\omega )=$ $\mathcal{A}_{\mathrm{ab/em}}(\omega )$.

In contrast to gauge violation spectroscopy, to calculate the gauge
invariant spectrum, one needs bound the matter field operator to the gauge
fields $\hat{\Psi}(\mathbf{r)\longrightarrow }\hat{\Psi}(\mathbf{r)}e^{i\int
d^{d}x\mathbf{E}_{\mathrm{cl}}(\mathbf{x})\cdot \mathbf{\hat{A}}(\mathbf{x}%
)} $, where $\mathbf{E}_{\mathrm{cl}}(\mathbf{x})$ is a classical electric
field satisfying $\nabla \cdot \mathbf{E}_{\mathrm{cl}}(\mathbf{x})=\delta
\left( \mathbf{x-r}\right) $. It is invariant under gauge transformation,
and commutes with the gauge charge $\left[ \hat{Q}(\mathbf{r}),\hat{\Psi}(%
\mathbf{r}^{\prime }\mathbf{)}e^{i\int d^{d}x\mathbf{E}_{\mathrm{cl}}(%
\mathbf{x})\cdot \mathbf{\hat{A}}(\mathbf{x})}\right] =0$. We note that in
most quantum simulators of synthetic gauge theories, it is hard to perform
the gauge invariant spectroscopy. Since it is challenging to excite such
highly non-local excitations. Therefore the commonly used techniques, such
as RF spectroscopy, probes the gauge violation spectrum rather than gauge
invariant spectrum. Furthermore, we will show that the gauge violation
spectroscopy can be used to detect the confinement and deconfinement phases
in synthetic gauge theories.

\begin{figure}[tbp]
\centering
\includegraphics[width=0.45\textwidth]{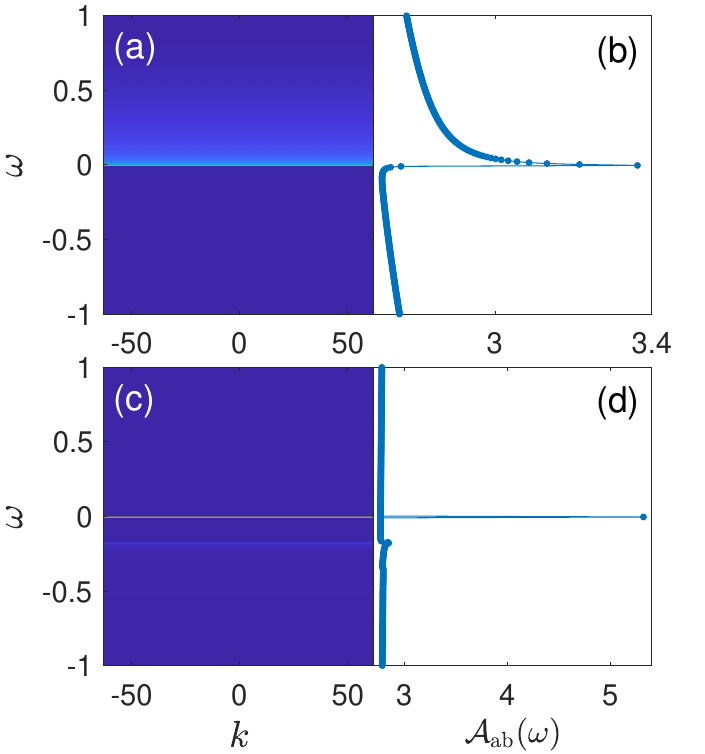}
\caption{Gauge violation single-particle absorbing spectrum for
deconfinement model (a)(b) and for confinement model (c)(d). (a)(c) Momentum
resolved gauge violation spectrum. (b)(d) Gauge violation spectrum at a
given momentum. We have set coupling constant $g=0.5$ for deconfinement
model, and the boson mass $m=e/\protect\sqrt{\protect\pi}=0.17$ for
confinement model.}
\label{f2Aab}
\end{figure}

\textit{Two Schwinger-like Models.} In the following, we will present two
one-dimensional models with U(1) local gauge symmetry, which can be both
exactly solved via the Bosonization method. One is the celebrated Schwinger
model. Its Hamiltonian is given by
\begin{equation}
\hat{H}^{\mathrm{c}}=\hat{H}_{\mathrm{mat}}+\frac{1}{2}\int dx\hat{E}^{2}(x),
\end{equation}%
where $\hat{H}_{\mathrm{mat}}=\int dx\hat{\Psi}^{\dag }(x)\sigma
_{z}[-i\partial _{x}-e\hat{A}(x)]\hat{\Psi}(x)$ and the fermion fields have
two components $\hat{\Psi}=(\hat{\Psi}_{R},\hat{\Psi}_{L})^{T}$. The
Schwinger model describes the (1+1) dimensional QED~\cite{Schwinger@1962.I,
Schwinger@1962.II}, and exhibits charge confinement phenomena. Another is similar
to the Schwinger model. Its Hamiltonian is given by%
\begin{equation}
\hat{H}^{\mathrm{d}}=\hat{H}_{\mathrm{mat}}+\frac{u}{2}\int dx\left[
\partial _{x}\hat{E}(x)\right] ^{2}.
\end{equation}%
Note that it possesses a modified Maxwell term. The corresponding energy
density of gauge field is proportional to the square of electrical field
gradient rather than the square of electrical field. This modified Maxwell
leads to the deconfinement of charges in this model.

These two models possess the local $U(1)$ gauge symmetries. The
corresponding gauge transformation operators are both $\hat{G}=\exp \left\{
i\int dx\left[ e\hat{\rho}(x)\theta (x)+\hat{E}(x)\partial _{x}\theta \right]
\right\} $, where $\theta (x)$ is an arbitrary phase distribution function,
and $\hat{\rho}(x)=\hat{\Psi}^{\dag }(x)\hat{\Psi}(x)$ is the particle
density operator. Since $\left[ \hat{G},\hat{H}^{\mathrm{c(d)}}\right] =0$,
the generator of this gauge transformation, $\hat{Q}(x)=\partial _{x}\hat{E}%
/e-\hat{\rho}(x)$ is a conserved quantity, which is also called gauge
charge. In the physical sector, $\hat{Q}(x)=0$, the conservation leads to
the Gaussian law in one dimension, $\partial _{x}\hat{E}=e\hat{\rho}(x)$.

We use the Bosonization method to deal with these two models. The Bosonization method maps one dimensional fermions to a problem of
bosonic fields~\cite{Giamarchi@2003.BK}. Here both $\hat{H}^{\mathrm{c}}$
and $\hat{H}^{\mathrm{d}}$ can be bosonized in arbitrary gauge sector. As
discussed above we only focus on the sector with gauge charge $\hat{Q}%
(x)=\xi \delta \left( x-x^{\prime }\right) $. The corresponding Bosonized
Hamiltonian in these sectors is given by,

\begin{eqnarray}
\hat{H}_{\xi }^{\mathrm{c}} &=&\hat{H}_{\mathrm{mat}}+\frac{m^{2}}{2}\int
dx\left( \phi +\frac{\xi e\mathrm{sgn}\left( x-x^{\prime }\right) }{2m}%
\right) ^{2}, \\
\hat{H}_{\xi }^{\mathrm{d}} &=&\hat{H}_{\mathrm{mat}}+\frac{g}{2}\int
dx\left( \partial _{x}\phi +\frac{\xi e\delta \left( x-x^{\prime }\right) }{m%
}\right) ^{2},
\end{eqnarray}%
where $\hat{H}_{\mathrm{mat}}=\frac{1}{2}\int dx\left[ \Pi ^{2}+\left(
\partial _{x}\phi \right) ^{2}\right] $, $g=um^{2}$ and $m^{2}=e^{2}/\pi $.
Note that all of these Hamiltonians are quadratic, thus can be exactly
diagonalized. Then it is straightforward to calculate the gauge violation
correlations defined in Eq. (\ref{GV.CR01},\ref{GV.CR02}). For the
deconfinement model, one obtains
\begin{eqnarray*}
ig^{>}(x,t) &=&\frac{\delta (x)}{2\pi a}\left( \frac{-ia}{vt-ia}\right)
^{\gamma }e^{-iE_{Q}t}, \\
ig^{<}(x,t) &=&\frac{\delta (x)}{2\pi a}\left( \frac{ia}{vt+ia}\right)
^{\gamma }e^{iE_{Q}t},
\end{eqnarray*}%
where $v=\sqrt{1+g}$ and $\gamma=\frac{v^2+1}{2v^3}$. The ultraviolet cutoff
$a$ is introduced by the bosonization procedure, and $E_{Q}$ is an
unimportant constant energy. For the Schwinger model, it is hard to obtain a
compact analytic form (see supplementary material). Then performing the Fourier
transformation one obtains the gauge violation spectral functions.

\begin{figure}[tbp]
\centering
\includegraphics[width=0.45\textwidth]{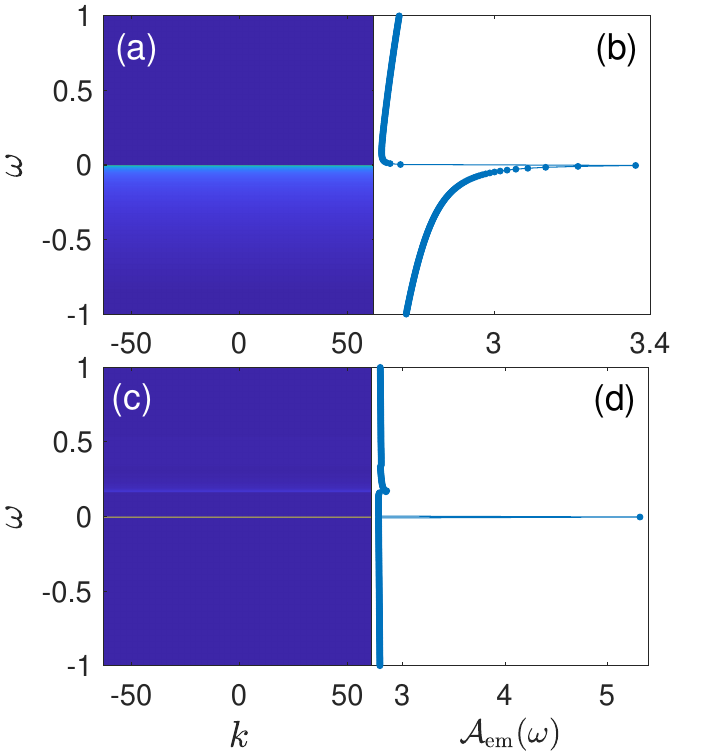}
\caption{Gauge violation single-particle emission spectrum for deconfinement
model (a)(b) and for confinement model (c)(d). (a)(c) Momentum resolved
gauge violation spectrum. (b)(d) Gauge violation spectrum at a given
momentum. We have set coupling constant $g=0.5$ for deconfinement model, and
the boson mass $m=e/\protect\sqrt{\protect\pi}=0.17$ for confinement model.}
\label{f3Aem}
\end{figure}

The results are shown in Figs.\ref{f2Aab} and \ref{f3Aem}. Note that for
both models, the gauge violation spectrums exhibit no dispersion. For the
model with charge confinement, the gauge violation spectrum is a delta
function. By contrast, for the deconfined model, the gauge violation
spectrum has a finite width. This behavior can be understood as follows:
The gauge violation spectroscopy measurement adds one particle with charge $%
+e$ into the system, add create a gauge charge $-e$ at the same position,
which can not move. The interaction between this particle and the gauge
charge is governed by electrical field. In the Schwinger model, this
interaction energy is proportional to the length of separation between the
added particle and the gauge charge, i.e. it is in the confinement phase.
Thus the added particle can not move far away from the original position.
However, in the deconfinement model, the interaction energy is nearly
constant. Then the added particle can move away from the original position.
For comparison, we also have calculated the gauge invariance spectroscopy in
supplementary material, which exhibit linear dispersion in momentum space.

\textit{Quantum Link Model. }Recently, the one-dimensional quantum link
model, a U(1) lattice gauge theory, has been realized on a quantum simulator
based on ultracold bosons in optical lattices~\cite{YZS@2020.QLM}. The Gauss law,
as well as the thermalization of this model have been observed~\cite{Yuan@2022.ETH}. By further engineering a tunable topological theta term,
the confinement-deconfinement transition has been observed~\cite{Yuan@2023.angle}. The
Hamiltonian of the quantum link model is given by~\cite{Zhai@2022.QLM,Jad@2023.QLM}
\begin{eqnarray}
\hat{H}_{\mathrm{QLM}} &=&J\sum_{j}\left( \hat{\Psi}_{j+1}^{\dag }\hat{S}%
_{j+1;j}^{-}\hat{\Psi}_{j}^{\dag }+h.c.\right)  \notag \\
&&+\sum_{j}M\hat{\Psi}_{j}^{\dag }\hat{\Psi}_{j}+\chi \sum_{j}(-1)^{j}\hat{S}%
_{j+1;j}^{z}.  \label{QLM}
\end{eqnarray}%
Here the gauge field is represented by spin-1/2 operators $\hat{S}%
_{i+1;i}^{+}$ on links. $J$ is the gauge-matter coupling strength and $M$
is the mass of the matter field. $\chi $ can tune the topological theta
angle $\theta $. When $\chi =0$, $\theta =\pi $, as $\chi \neq 0$, $\theta $
is tuned away from $\pi $. When $\chi \neq 0$, i.e. $\theta $ is away from $%
\pi $, it is in the confined phase, and no single charge can be observed.
When $\chi =0$, i.e. $\theta =\pi $, there is a transition from confined
phase to deconfined phase by tuning the matter mass from negative to
positive.

\begin{figure}[t]
\includegraphics[width=0.45\textwidth]{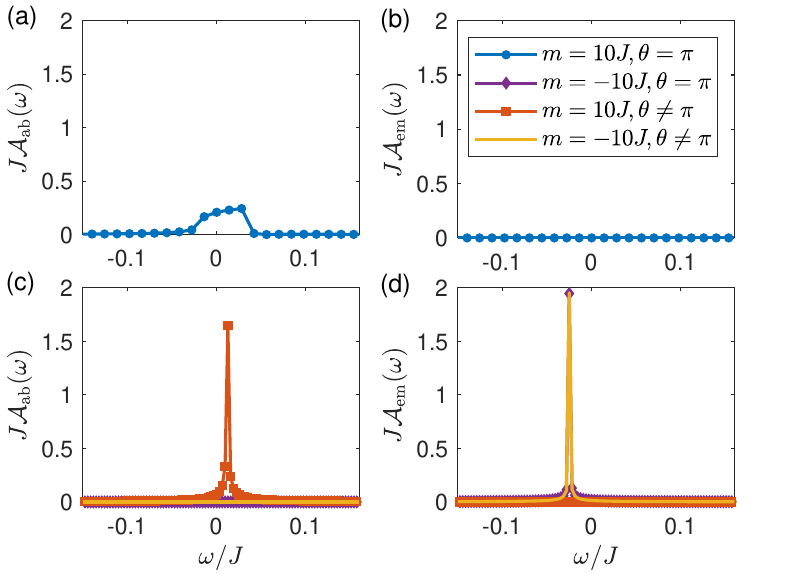}
\caption{Gauge violation spectroscopies for the one-dimensional quantum link
model. (a)(b) Results for deconfinement phase with $M=10J,\protect\theta =%
\protect\pi $. (c)(d) Results for confinement phase with $M=-10J,\protect%
\theta =\protect\pi $ and $M=\pm10J,\protect\theta =\protect\pi /2$. The
model contains total number of sites $L=20$.}
\label{f4QLM}
\end{figure}

The local gauge transformation operator is $\hat{G}_{j}=e^{i\phi \hat{Q}%
_{j}} $, and the conserved gauge charge is $\hat{Q}_{j}=\hat{S}_{j+1;j}^{z}+%
\hat{S}_{j;j-1}^{z}+\hat{\Psi}_{j}^{\dag }\hat{\Psi}_{j}$. We calculate the
gauge violation spectrum by numerical exact diagonalization. The results of
numerical simulation are shown in Fig.\ref{f4QLM}. In Figs.\ref{f4QLM}(a-b),
we observe the deconfinement behavior with $M=10J$ and the topological angle
$\theta =\pi $. The spectrum has a finite width. Instead, Fig.\ref{f4QLM}%
(c-d) shows the confinement behavior with $M=-10J,\theta =\pi $. It shows
the delta function behavior. This difference clearly distinguishes
confinement from deconfinement phase. In Fig.\ref{f4QLM}(c-d), we also
observe the same delta function behavior with $M=\pm 10J,\theta \neq \pi $,
because there is only a confinement phase when $\theta \neq \pi $.

\textit{Summary.} We propose the gauge violation spectroscopy in synthetic
gauge theories on quantum simulators, which could be used to detect the
confinement phase or deconfinement phase. In most single-particle
spectroscopy of synthetic gauge theories, such as RF spectroscopy in ultracold
quantum gases, one measured the gauge violation spectrum, rather than the
gauge invariant spectrum. Since later spectroscopy needs highly non-local
perturbations. We used three one-dimensional models with local U(1) gauge
symmetry to show that in the confined phase, gauge violation spectrum is
nearly a delta-function, while in the deconfinement phase the spectrum has a
finite width. However, our conclusions are not limited to one-dimensional
models with U(1) gauge symmetry. It can be applied to higher dimensions or non-Abelian gauge.
In addition, we would like to point out that, for some simulators, people realize the projected Hamiltonian instead of the Hamiltonian in the full Hilbert space~\cite{Surace@2020.LGT}. In this situation, there is no room for fluctuation of gauge charge. Thus the gauge violation spectroscopy can not be applied in these simulators.

\textit{Acknowledgements.} We thank Hui Zhai and Yanting Cheng for
discussion. This work is supported by Innovation Program for Quantum Science
and Technology (Grant No. 2021ZD0302000).


\begin{widetext}
\section{Schwinger-Like Models}
\label{model and derivation}
In this supplementary material, we begin by reviewing the Schwinger model with massless fermion, which contains a confinement phase, and presenting a deconfinement model that is similar to the Schwinger model but possesses a modified Maxwell term.

\textbf{Schwinger-Like Models.}
In the following we present two one-dimensional models with U(1) local gauge symmetry, the celebrated Schwinger model and deconfinement model with a modified Maxwell term denoted by $\hat{H}^{\mathrm{c}},\hat{H}^{\mathrm{d}}$ respectively,
\begin{eqnarray}
\hat{H}^{\mathrm{c}}&=&\hat{H}_{\mathrm{mat}}+\frac{1}{2}\int dx\hat{E}^{2}(x),\label{conf}\\
\hat{H}^{\mathrm{d}}&=&\hat{H}_{\mathrm{mat}}+\frac{u}{2}\int dx\left[\partial _{x}\hat{E}(x)\right] ^{2}.\label{deconf}
\end{eqnarray}
where $\hat{H}_{\mathrm{mat}}=\int dx\hat{\Psi}^{\dag }(x)\sigma
_{z}[-i\partial _{x}-e\hat{A}(x)]\hat{\Psi}(x)$ and the fermion fields have
two components $\hat{\Psi}=(\hat{\Psi}_{R},\hat{\Psi}_{L})^{T}$. The (anti-)commutation relations are
\begin{equation}\label{}
\begin{aligned}
\left\{\hat{\Psi}_{\alpha}(x), \hat{\Psi}_{\beta}^{\dagger}(y)\right\} &=\delta_{\alpha \beta} \delta(x-y), \\
{\left[\hat{A}(x), \hat{E}(y)\right] } &=-i \delta(x-y),
\end{aligned}
\end{equation}
with all other (anti-)commutators vanishing.

These two models possess the local U(1) gauge symmetries. Quantum mechanically, the gauge transformation of the two models is implemented by the same unitary operator
\begin{equation}\label{unitary operator}
\begin{aligned}
  \hat{G}=&\exp \left\{i \int d x\left[e\hat{\rho}(x) \theta(x)+\hat{E}(x) \partial_{x} \theta(x)\right]\right\}\\
  =&\exp \left\{i e\int d x\left[\hat{\rho}(x) - \frac{1}{e}\partial_{x}\hat{E}(x)\right]\theta(x)\right\},
\end{aligned}
\end{equation}
that act as $\hat{G} \hat{\Psi}(x) \hat{G}^{\dag} =\hat{\Psi}(x) e^{-i e \theta(x)}$ and $\hat{G}(x) \hat{A}(x) \hat{G}^{\dag} =\hat{A}(x)-\partial_{x} \theta(x),$ where $\theta(x)$ is an arbitrary phase distribution function, and $\hat{\rho}(x)=\hat{\Psi}^{\dag}(x)\hat{\Psi}(x)$ is the particle density operator. Since the two Hamiltonians \eqref{conf}\eqref{deconf} are invariant under the gauge transformation, i.e., $\hat{G}\hat{H}^{\mathrm{c(d)}}\hat{G}^{\dag}=\hat{H}^{\mathrm{c(d)}},$ we can conclude that
\begin{equation}\label{Gauss's law}
  \partial_{x} \hat{E}(x)=e\left(\hat{\rho}(x)+\hat{Q}(x)\right),
\end{equation}
where the conserved quantity $\hat{Q}(x)$ is local gauge charge separating the Hamiltonian into different sectors. A physically meaningful state must be invariant to the local gauge transformation and therefore must be annihilated by the operator $\hat{Q}(x)$
\begin{equation}\label{Gl}
  \hat{Q}(x)|\text{physical state}\rangle=0,
\end{equation}
which is nothing but the Gauss's law analogous to that in three dimension.

For simplicity, we fix a particular gauge such that $\hat{A}(x)$ vanishes. For the purpose, we choose $\partial_x \theta(x)=\hat{A}(x)$, then the unitary operator becomes
\begin{equation}\label{}
\begin{aligned}
\hat{G} &=\exp \left\{i \int d x\left[e\hat{\rho}(x) \int_{-\infty}^{x} \hat{A}(y) d y+\hat{E}(x) \hat{A}(x)\right]\right\} \\
&=\exp\left\{i \int d x\left[\hat{A}(x) \int_{x}^{+\infty} e \hat{\rho}(y)dy+\hat{E}(x) \hat{A}(x)\right]\right\}
\end{aligned}
\end{equation}
that act as $\hat{G} \hat{\Psi}(x) \hat{G}^{\dag}=\hat{\Psi}(x) e^{-i e \int_{-\infty}^{x} d y \hat{A}(y)}, \hat{G} \hat{E}(x) \hat{G}^{\dagger} =2 \hat{E}(x)+e \int_{x}^{+\infty}\hat{\rho}(y)dy$ and $\hat{G} \hat{A}(x) \hat{G}^{\dag}=0,$ furthermore,
\begin{equation}\label{}
\begin{aligned}
  \hat{G}\partial_x \hat{E}(x) \hat{G}^{\dag}&=2 \partial_x \hat{E}(x)-e \hat{\rho}(x),\\
  \hat{G} \hat{Q}(x) \hat{G}^{\dagger}&=2\hat{Q}(x).
\end{aligned}
\end{equation}
In the gauge, we arrive at the simplified Hamiltonian
\begin{eqnarray}
\hat{H}^{\mathrm{c}}&=&\int dx\hat{\Psi}^{\dag }(x)\sigma
_{z}[-i\partial _{x}]\hat{\Psi}(x)+\frac{1}{2}\int dx\hat{E}^{2}(x),\label{gaugefixedconH}\\
\hat{H}^{\mathrm{d}}&=&\int dx\hat{\Psi}^{\dag }(x)\sigma
_{z}[-i\partial _{x}]\hat{\Psi}(x)+\frac{u}{2}\int dx\left[\partial _{x}\hat{E}(x)\right] ^{2}.\label{gaugefixeddeconH}
\end{eqnarray}
Obviously, the relations $\partial_x \hat{E}(x)=e\left(\hat{Q}(x)+\hat{\rho}(x)\right)$ and $[\hat{Q}(x),\hat{H}^{\mathrm{c(d)}}]=0$ are still satisfied by the more compact Hamiltonian.

\section{Bosonization Method}
To calculate the spectral functions, we work out the bosonic version of the Schwinger-like models Eq.\eqref{gaugefixedconH}\eqref{gaugefixeddeconH}, which the Bogoliubov transformation can diagonalize in any sector. Before bosonizing the two models, the bosonization dictionary we will use in the following is given,
\begin{gather}\label{}
\hat{\Psi}_{R(L)}(x)=\frac{1}{\sqrt{2\pi a}}e^{(-)i\sqrt{4\pi}\hat{\phi}_{R(L)}(x)},\label{FB}\\
\hat{\rho}(x)=\hat{\Psi}_R^{\dag}\hat{\Psi}_R+\hat{\Psi}_L^{\dag}\hat{\Psi}_L=\frac{1}{\sqrt{\pi}} \partial_{x} \hat{\phi}\label{rho},\\
\hat{H}_{free} =\int_{-L / 2}^{L / 2} d x\left[\hat{\Psi}_{R}^{\dag}(x)\left(-i \partial_{x}\right) \hat{\Psi}_{R}(x)+\hat{\Psi}_{L}^{\dag}(x)\left(i \partial_{x}\right) \hat{\Psi}_{L}(x)\right]\notag\\
=\frac{1}{2} \int_{-L / 2}^{L / 2} d x\left[\hat{\Pi}^2+\left(\partial_x \hat{\phi}\right)^2\right],\label{Hfree}
\end{gather}
and some definitions in the above equations
\begin{gather}\label{bosondef}
  \hat{\phi}_{R}(x)=\frac{1}{\sqrt{2L}}\sum_{q}\Theta(+q)\frac{1}{|q|}\left(\hat{b}_q e^{iqx}+\hat{b}^\dag_q e^{-iqx}\right),\notag\\
  \hat{\phi}_{L}(x)=\frac{1}{\sqrt{2L}}\sum_{q}\Theta(-q)\frac{1}{|q|}\left(\hat{b}_q e^{iqx}+\hat{b}^\dag_q e^{-iqx}\right),\notag\\
  \hat{\phi} =\hat{\phi}_{R}+\hat{\phi}_{L},\label{bosondef}\\
  \hat{\theta} =\hat{\phi}_{L}-\hat{\phi}_{R},\notag\\
  \hat{\Pi}(x)=\partial_{x} \hat{\theta}(x).\notag
\end{gather}
where $\hat{b}_{q}$ is the usual bosonic annihilation operator in the state with momentum $q$ satisfying the commutation relations $[\hat{b}_{q},\hat{b}_{q^{\prime}}^\dag]=\delta_{qq^{\prime}}$. $e^{-a q / 2}$ is a converging factor with $a\rightarrow0^+$. The system size $L$ will be set to infinity in the last step.

\textbf{Bosonization.}
It is straightforward to bosonize the first term in the two models Eq.\eqref{gaugefixedconH}\eqref{gaugefixeddeconH} using the bosonization dictionary Eq.\eqref{Hfree}. Thus, let's focus on the second term, which includes the electric field. We can work out the electric field $\hat{E}(x)$ from Gauss’s law \eqref{Gauss's law} using Green's function method,
\begin{align}\label{}
  \hat{E}(x)=\int d y \frac{e}{2}\{\Theta(x-y)-\Theta(y-x)\}(\hat{Q}(y)+\hat{\rho}(x)) d y.
\end{align}
Then use the bosonic expression of $\hat{\rho}(x)$ Eq.\eqref{rho}, and straightforward integration leads to
\begin{align}\label{elecF}
  \hat{E}(x)=\hat{E}_{B}(x)+\frac{e}{\sqrt{\pi}}\hat{\phi}(x),
\end{align}
where the background electric field generated by the gauge charge $Q(x)$ is defined as
\begin{align}\label{EB}
    \hat{E}_{B}(x)=\int d y \frac{e}{2} \operatorname{sgn}(x-y) \hat{Q}(y) d y.
\end{align}
Take the spatial derivative on both sides of Eq.\eqref{elecF}, and we have
\begin{align}\label{pelecF}
  \partial_x \hat{E}(x)=e \hat{Q}(x)+\frac{e}{\sqrt{\pi}}\partial_x\hat{\phi}(x).
\end{align}
Substitute Eq.\eqref{elecF} and \eqref{pelecF} into Eq.\eqref{gaugefixedconH} and \eqref{gaugefixeddeconH} respectively, and we arrive at the bosonized Hamiltonians in any sector $\hat{Q}(x)$ for the Schwinger-like models,
\begin{eqnarray}
  \hat{H}^{\mathrm{c}}&=&\int dx\frac{1}{2}\left[\hat{\Pi}^{2}+(\partial_x \hat{\phi})^{2}+m^2\left(\frac{\sqrt{\pi}}{e} \hat{E}_{B}(x)+\hat{\phi}(x)\right)^{2}\right],\\
  \hat{H}^{\mathrm{d}}&=&\int dx\frac{1}{2}\left[\hat{\Pi}^{2}+(\partial_x \hat{\phi})^{2}+g\left(\sqrt{\pi}\hat{Q}(x)+\partial_x\hat{\phi} \right)^{2}\right]
\end{eqnarray}\label{boseHamiltonian}
with the mass of the bosonic field $\hat{\phi}$, $m^2=\frac{e^{2}}{\pi}$, and a dimensionless coupling constant $g=ue^2/\pi$. In sectors $\hat{Q}(x)=\xi\delta(x-x^\prime)$ with $\xi=0,\pm$, we obtain $\hat{H}^{\mathrm{c(d)}}_\xi$ in the main text.

\section{Spectral functions}
In the following, we want to calculate the following types of correlations
\begin{eqnarray}
ig_\sigma^{>}(x, t)&=&\left\langle\psi_{0}\left|\hat{\Psi}_{R}(x) e^{-i \hat{H}_{\xi}^\sigma t} \hat{\Psi}_{R}^{\dagger}(0)\right| \psi_{0}\right\rangle e^{+i E_{0}^\sigma t}, \label{G>}\\
ig_\sigma^{<}(x, t)&=&\left\langle\psi_{0}\left|\hat{\Psi}_{R}^{\dagger}(0) e^{+i \hat{H}_{\xi}^\sigma t} \hat{\Psi}_{R}(x)\right| \psi_{0}\right\rangle e^{-i E_{0}^\sigma t}.\label{G<}
\end{eqnarray}
Here, $\hat{H}_{\xi}^\sigma$ with $\sigma=\mathrm{c,d}$ denote the Hamiltonians of Schwinger-like models as in the main text or see Eq.\eqref{BConfH}\eqref{BDconfH}. The two models can be discussed in a fully parallel and unified fashion by introducing the parameter $\sigma$. The $\left\vert\psi_0\right\rangle$ and $E_0^\sigma$ are the ground state and energy in the physical sector, $\hat{H}_{\xi=0}^\sigma\left\vert\psi_0\right\rangle=E_0^\sigma\left\vert\psi_0\right\rangle$. Since the right and left movers possess the same physics as the other, we only focus on the right mover in the following.

Firstly, we diagonalize the bosonic Schwinger-like models in sectors $\hat{Q}(x)=\xi\delta(x)$ with $\xi=0,\pm$,
\begin{eqnarray}
\hat{H}_{\xi }^{\mathrm{d}} &=&\hat{H}_{\mathrm{mat}}+\frac{m^{2}}{2}\int
dx\left( \phi +\frac{\xi e\mathrm{sgn}\left( x \right) }{2m}%
\right) ^{2}, \label{BConfH}\\
\hat{H}_{\xi }^{\mathrm{c}} &=&\hat{H}_{\mathrm{mat}}+\frac{g}{2}\int
dx\left( \partial _{x}\phi +\frac{\xi e\delta \left( x \right) }{m%
}\right) ^{2}.\label{BDconfH}
\end{eqnarray}%
It is important to note that the definition of the bosonic field Eq\eqref{bosondef} and the Fourier expansion of $\operatorname{sgn}\left(x\right) =\frac{1}{L} \sum_{k} \frac{2}{i k} e^{i k x}$ allow us to rewrite the Hamiltonians into a unified form in momentum space
\begin{equation}\label{}
  \hat{H}_{\xi}^{\sigma}=E_{\mathrm{zp}}^{\sigma}+\xi^{2} E_{Q}^{\sigma}+\sum_{q}\left\{\mu_{q}^{\sigma} \hat{b}_{q}^{\dagger} \hat{b}_{q}+\frac{1}{2} \Omega_{q}^{\sigma}\left(\hat{b}_{q} \hat{b}_{-q}+\hat{b}_{-q}^{\dagger} \hat{b}_{q}^{\dagger}\right)+i \xi \lambda_{q}^{\sigma}\left(\hat{b}_{q}-\hat{b}_{q}^{\dagger}\right)\right\}.
\end{equation}
Here,
\begingroup
\renewcommand*{\arraystretch}{1.5}
\begin{equation}
\begin{gathered}
\left\{\begin{array}{l}
E_{\mathrm{zp}}^{\mathrm{c}}=\sum_q \frac{1}{2}\left(|q|+\frac{m^2}{2|q|}\right) \\
E_{\mathrm{zp}}^{\mathrm{d}}=\sum_q \frac{1}{2}\left(1+\frac{g}{2}\right)|q|
\end{array},\right. \quad
\left\{\begin{array}{l}
E_Q^{\mathrm{c}}=\sum_q \frac{e^2}{2 L q^2} \\
E_Q^{\mathrm{d}}=\sum_q \frac{e^2 g}{2 L m^2}
\end{array},\right. \\
\left\{\begin{array}{l}
\mu_q^{\mathrm{c}}=|q|+\frac{m^2}{2|q|} \\
\mu_q^{\mathrm{d}}=(1+\frac{g}{2})|q|
\end{array},\right. \quad
\left\{\begin{array}{l}
\Omega_q^{\mathrm{c}}=\frac{m^2}{2|q|} \\
\Omega_q^{\mathrm{d}}=\frac{g|q|}{2}
\end{array},\right.\quad
\left\{\begin{array}{l}
\lambda_q^{\mathrm{c}}=\frac{e m}{q \sqrt{2 L|q|}}\\
\lambda_q^{\mathrm{d}}=\frac{e g q}{m \sqrt{2 L|q|}}
\end{array}.\right.
\end{gathered}
\end{equation}
\endgroup
So the Hamiltonian in the physical sector is given by
\begin{equation}\label{}
  \hat{H}_{\xi=0}^{\sigma}=E_{\mathrm{zp}}^{\sigma}+\sum_{q}\left\{\mu_{q}^{\sigma} \hat{b}_{q}^{\dagger} \hat{b}_{q}+\frac{1}{2} \Omega_{q}^{\sigma}\left(\hat{b}_{q} \hat{b}_{-q}+\hat{b}_{-q}^{\dagger} \hat{b}_{q}^{\dagger}\right)\right\},
\end{equation}
which can be diagonalized by utilizing the Bogoliubov transformation
\begin{eqnarray}
\hat{b}_q&=&u_q^\sigma \hat{\beta}_q-v_q^\sigma \hat{\beta}_{-q}^{\dagger}, \\
\hat{b}_q^{\dagger}&=&u_q^\sigma \hat{\beta}_q^{\dagger}-v_q^\sigma \hat{\beta}_{-q},
\end{eqnarray}
with
\begingroup
\renewcommand*{\arraystretch}{1.5}
\begin{equation}
\left\{\begin{array}{l}
\left(u_q^\sigma\right)^2=\frac{1}{2}\left(\frac{\mu_q^\sigma}{\varepsilon_q^\sigma}+1\right) \\
\left(v_q^\sigma\right)^2=\frac{1}{2}\left(\frac{\mu_q^\sigma}{\varepsilon_q^\sigma}-1\right)
\end{array}\right.,\quad
\left\{\begin{array}{l}
\varepsilon_q^{\mathrm{c}}=\sqrt{|q|^2+m^2} \\
\varepsilon_q^{\mathrm{d}}=\sqrt{1+g}|q|
\end{array}.\right. \\
\end{equation}
\endgroup
It follows that
\begin{equation}\label{physicalH}
  \hat{H}_{\xi=0}^\sigma=E_0^\sigma+\sum_{q}\varepsilon_q^\sigma\hat{\beta}^\dag_q\hat{\beta}_q,
\end{equation}
where $E_0^\sigma=E_{\mathrm{zp}}^\sigma+E_{\mathrm{LHY}}^\sigma$ is the ground state energy in the physical sector, and $E_{\mathrm{LHY}}=\frac{1}{2}\sum_{q}(\varepsilon_q^\sigma-\mu_q^\sigma)$.
In this basis, the Hamiltonians in sectors $\hat{Q}(x)=\xi\delta(x)$ can be rewritten into
\begin{equation}\label{nonphysicalH}
\begin{aligned}
\hat{H}_{\xi}^\sigma & =E_0^\sigma+\xi^2 E_Q^\sigma+\sum_q \varepsilon_q^\sigma \hat{\beta}_q^{\dagger} \hat{\beta}_q+\sum_q i \xi \lambda_q^\sigma\left(u_q^\sigma-v_q^\sigma\right)\left(\hat{\beta}_q-\hat{\beta}_q^{\dagger}\right) \\
& =E_0^\sigma+\xi^2 \tilde{E}_Q^\sigma+\sum_q \varepsilon_q^\sigma\left(\hat{\beta}_q^{\dagger}+\xi h_q^{\sigma *}\right)\left(\hat{\beta}_q+\xi h_q^\sigma\right)
\end{aligned}
\end{equation}
with
\begin{eqnarray}
h_q^\sigma&=&-i \frac{\lambda_q^\sigma}{\varepsilon_q^\sigma}\left(u_q^\sigma-v_q^\sigma\right), \\
\tilde{E}_Q^\sigma&=&E_Q^\sigma-\sum_q\left(\lambda_q^\sigma\right)^2\left(u_q^\sigma-v_q^\sigma\right)^2 / \varepsilon_q^\sigma.
\end{eqnarray}

To calculate spectral functions, it is also convenient to rewrite the bosonic fields Eq.\eqref{FB} in terms of the new basis
\begin{equation}\label{FB2}
\hat{\Psi}_R(x)=\frac{1}{\sqrt{2 \pi a}} e^{\sum_q\left[\eta_q^\sigma(x) \hat{\beta}_q^{\dagger}-\eta_q^{\sigma *}(x) \hat{\beta}_q\right]} \hat{F}_{\mathrm{R}}
\end{equation}
with
\begin{equation}
\eta_q^\sigma(x)=i \sqrt{\frac{2 \pi}{|q| L}}\left[\Theta(+q) u_q^\sigma-\Theta(-q) v_q^\sigma\right] e^{-i q x}.
\end{equation}

\textbf{Gauge Invariance Spectroscopy.}
Substitute Eq.\eqref{physicalH}\eqref{FB2} into the Green's function Eq.\eqref{G>}, and we obtain
\begin{equation}\label{}
  ig_\sigma^>(x,t)=\frac{1}{2\pi a}\left\langle\psi_0\right| e^{+\sum_{q}\left[ \eta_q^\sigma(x)\hat{\beta}_q^\dag
  - \eta_q^{*\sigma}(x)\hat{\beta}_q\right]} e^{-it\sum_{q}\varepsilon_q^\sigma\hat{\beta}_q^\dag\hat{\beta}_q} e^{-\sum_{q}\left[ \eta_q^\sigma(0)\hat{\beta}_q^\dag - \eta_q^{*\sigma}(0)\hat{\beta}_q\right]}\left|\psi_0\right\rangle,
\end{equation}
and $ig_\sigma^<(x,t)$ is similar. Since we are calculating the correlations for the ground state $\left\vert\psi_0\right\rangle$ of $\hat{H}_{\xi=0}$, we can work out the expression by normal-ordering it using Baker-Hausdorff (BH) formula. For deconfinement model, straightforward algebra gives
\begin{eqnarray}
  ig_\mathrm{d}^>(x,t)=\frac{a^{\frac{(v-1)^2}{2v}}}{2\pi}\left(i(x+vt)+a\right)^{-\frac{(v+1)^2}{4v}}
  \left(i(-x+vt)+a\right)^{-\frac{(v-1)^2}{4v}},\\
  ig_\mathrm{d}^<(x,t)=\frac{a^{\frac{(v-1)^2}{2v}}}{2\pi}\left(i(-x-vt)+a\right)^{-\frac{(v+1)^2}{4v}}
  \left(i(x-vt)+a\right)^{-\frac{(v-1)^2}{4v}}.
\end{eqnarray}
where $v=\sqrt{1+g}$ with dimensionless coupling constant $g=ue^2/\pi$. For confinement model, it gives
\begin{eqnarray}
  ig_\mathrm{c}^>(x,t)&=&\frac{1}{2\pi a}\exp\left\{-\int_{2\pi/L}^{\infty} dq e^{-a q} \frac{m^{4}}{4 q^{2} \varepsilon_{q}^\mathrm{c}}\left[\frac{1-e^{-i (q x+\varepsilon_{q}^\mathrm{c}t)}}{\left(\varepsilon_{q}^\mathrm{c}+q\right)^{2}}+\frac{1-e^{i (q x-\varepsilon_{q}^\mathrm{c}t)}}{\left(\varepsilon_{q}^\mathrm{c}-q\right)^{2}}\right]\right\},\\
  ig_\mathrm{c}^<(x,t)&=&\frac{1}{2\pi a}\exp\left\{-\int_{2\pi/L}^{\infty} dq e^{-a q} \frac{m^{4}}{4 q^{2} \varepsilon_{q}^\mathrm{c}}\left[\frac{1-e^{i (q x+\varepsilon_{q}^\mathrm{c}t)}}{\left(\varepsilon_{q}^\mathrm{c}+q\right)^{2}}+\frac{1-e^{-i (q x-\varepsilon_{q}^\mathrm{c}t)}}{\left(\varepsilon_{q}^\mathrm{c}-q\right)^{2}}\right]\right\}.
\end{eqnarray}
Then we can calculate the spectral functions using the Fourier transformation. The results are shown in Fig.\ref{inv}, which exhibit linear dispersion in momentum space.
\begin{figure}
  \centering
  \includegraphics[width=0.8\textwidth]{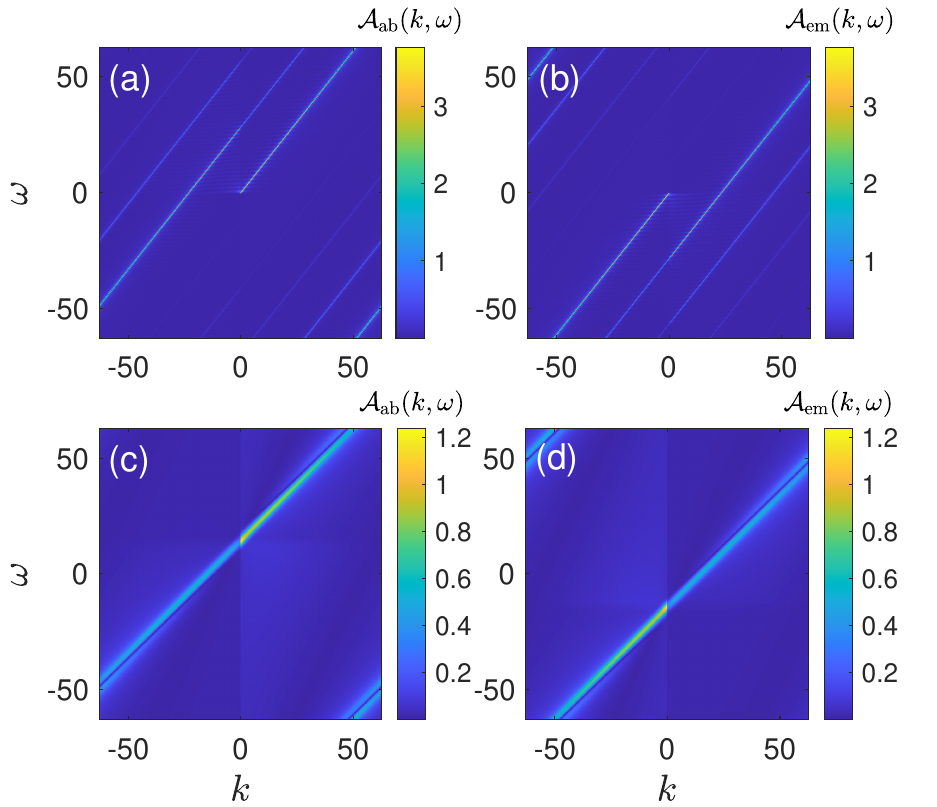}
  \caption{The gauge invariance single-particle absorbing and emission spectrum for deconfinement Model (a)(b) and confinement Model (c)(d). We have set coupling constant $g=0.5$, the mass of boson $m=e/\sqrt{\pi}=0.17$, and system size $L\rightarrow+\infty$.}\label{inv}
\end{figure}

\textbf{Gauge Violation Spectroscopy.}
Because of the orthogonality of states in different gauge sectors, the gauge violation correlations can be written as
\begin{eqnarray}
ig_\sigma^{>}(x, t)&=&\delta(x)\left\langle\psi_{0}\left|\hat{\Psi}_{R}(0) e^{-i \hat{H}_{\xi}^\sigma t} \hat{\Psi}_{R}^{\dagger}(0)\right| \psi_{0}\right\rangle e^{+i E_{0}^\sigma t},\\
ig_\sigma^{<}(x, t)&=&\delta(x)\left\langle\psi_{0}\left|\hat{\Psi}_{R}^{\dagger}(0) e^{+i \hat{H}_{\xi}^\sigma t} \hat{\Psi}_{R}(0)\right| \psi_{0}\right\rangle e^{-i E_{0}^\sigma t}.
\end{eqnarray}
Substitute Eq.\eqref{nonphysicalH}\eqref{FB2} into the above correlations, and we obtain
\begin{equation}\label{}
\begin{aligned}
  ig_\sigma^>(x,t)=\frac{\delta(x)}{2\pi a}&\left\langle\psi_0\right| e^{+\sum_{q}\left[ \eta_q^\sigma(0)\hat{\beta}_q^\dag
  - \eta_q^{*\sigma}(0)\hat{\beta}_q\right]} e^{-it\sum_{q}\varepsilon_q^\sigma\left(\hat{\beta}_q^\dag+\xi h_q^{\sigma *}\right)\left(\hat{\beta}_q+\xi h_q^{\sigma }\right)} \\
  &\times e^{-\sum_{q}\left[ \eta_q^\sigma(0)\hat{\beta}_q^\dag - \eta_q^{*\sigma}(0)\hat{\beta}_q\right]}\left|\psi_0\right\rangle e^{-i\xi^2\tilde{E}_Q^\sigma t},
\end{aligned}
\end{equation}
and $ig_\sigma^<(x,t)$ is similar. Note that $e^{+\sum_{q}\left[ \eta_q^\sigma(0)\hat{\beta}_q^\dag - \eta_q^{*\sigma}(0)\hat{\beta}_q\right]}$ is a translation operator
\begin{equation}
\begin{aligned}
& e^{+\sum_q\left[\eta_q^\sigma(0) \hat{\beta}_q^{\dagger}-\eta_q^{\sigma *}(0) \hat{\beta}_q\right]} \hat{\beta}_q^{\dagger} e^{-\sum_q\left[\eta_q^\sigma(0) \hat{\beta}_q^{\dagger}-\eta_q^{\sigma *}(0) \hat{\beta}_q\right]}=\hat{\beta}_q^{\dagger}-\eta_q^{\sigma *}, \\
& e^{+\sum_q\left[\eta_q^\sigma(0) \hat{\beta}_q^{\dagger}-\eta_q^{\sigma *}(0) \hat{\beta}_q\right]} \hat{\beta}_q e^{-\sum_q\left[\eta_q^\sigma(0) \hat{\beta}_q^{\dagger}-\eta_q^{\sigma *}(0) \hat{\beta}_q\right]}=\hat{\beta}_q-\eta_q^\sigma.
\end{aligned}
\end{equation}
Thus we obtain
\begin{equation}\label{}
  ig_\sigma^>(x,t)=\frac{\delta(x)}{2\pi a}\left\langle\psi_0\right| e^{-it\sum_{q}\varepsilon_q^\sigma\left(\hat{\beta}_q^\dag+\chi_q^{\sigma *}\right)\left(\hat{\beta}_q+\chi_q^{\sigma }\right)}\left|\psi_0\right\rangle e^{-i\xi^2\tilde{E}_Q^\sigma t}
\end{equation}
with $\chi_q^\sigma=\xi h_q^\sigma-\eta_q^\sigma(0)$. Then, we can define $\hat{B}_q^\sigma=\hat{\beta}_q+\chi_q^{\sigma }$, such that $\hat{B}_q^\sigma\left\vert\psi_0\right\rangle=\chi_q^{\sigma }\left\vert\psi_0\right\rangle$. That is to say the ground state $\left\vert\psi_0\right\rangle$ in physical sector is the coherent state of operator $\hat{B}_q^\sigma$. Thus we have
\begin{equation}
\begin{aligned}
ig_\sigma^>(x,t)& =\frac{\delta(x)}{2\pi a}\left\langle\psi_0\left|e^{-i t \sum_q \varepsilon_q^\sigma \hat{B}_q^{\dagger} \hat{B}_q}\right| \psi_0\right\rangle e^{-i \xi^2 \tilde{E}_Q^\sigma t} \\
& =\exp \left\{-i \xi^2 \tilde{E}_Q^\sigma t+\sum_q\left|\chi_q^\sigma\right|^2\left(e^{-i \varepsilon_q^\sigma t}-1\right)\right\}
\end{aligned}
\end{equation}
Now let us calculate the summation
\begin{equation}\label{}
  \sum_q\left|\chi_q^\sigma\right|^2\left(e^{-i \varepsilon_q^\sigma t}-1\right)=\sum_{q>0}\frac{2\pi}{|q|L}\gamma_{q,\sigma}^{+}\left(e^{-i \varepsilon_q^\sigma t}-1\right)+
  \sum_{q<0}\frac{2\pi}{|q|L}\gamma_{q,\sigma}^{-}\left(e^{-i \varepsilon_q^\sigma t}-1\right),
\end{equation}
where we have defined
\begin{eqnarray}
  \gamma_{q,\sigma}^{+}&=&\left|1+\xi\sqrt{\frac{|q|L}{2\pi}}\frac{\lambda_q^\sigma}{\varepsilon_q^\sigma}(u_q^\sigma-v_q^\sigma)\right|^2,\\
  \gamma_{q,\sigma}^{-}&=&\left|1+\xi\sqrt{\frac{|q|L}{2\pi}}\frac{\lambda_q^\sigma}{\varepsilon_q^\sigma}(v_q^\sigma-u_q^\sigma)\right|^2.
\end{eqnarray}

For the deconfinement model, $\gamma_{q,\mathrm{d}}^{+,-}$ is independent of $q$ because of the linearity of the dispersion relation $\varepsilon_q^d=\sqrt{1+g}|q|$. Thus, we can calculate the summation by using the formula $\ln(1-x)=-\sum_{n>0}x^n/n$, and the gauge violation correlations read
\begin{eqnarray}
ig_\mathrm{d}^>(x,t)&=&\frac{\delta(x)}{2\pi a}(\frac{-ia}{vt-ia})^{\frac{v^2+1}{2v^3}},\\
ig_\mathrm{d}^<(x,t)&=&\frac{\delta(x)}{2\pi a}(\frac{ia}{vt+ia})^{\frac{v^2+1}{2v^3}}.
\end{eqnarray}

For the confinement model, since $\gamma_{q,\mathrm{c}}^{+,-}$ depends on $q$, it is hard to obtain a compact analytic form. Here, we only show the exponential integral
\begin{align}
\begin{aligned}
    ig_\mathrm{c}^>(x,t)=\frac{\delta(x)}{2\pi a}&\exp\left\{-\int_{2\pi/L}^{\infty} dq e^{-a q} \frac{m^{4}}{4 q^{2} \varepsilon_{q}^\mathrm{c}}\left[\frac{1-e^{-i \varepsilon_{q}^\mathrm{c}t}}{\left(\varepsilon_{q}^\mathrm{c}+q\right)^{2}}+\frac{1-e^{-i \varepsilon_{q}^\mathrm{c}t}}{\left(\varepsilon_{q}^\mathrm{c}-q\right)^{2}}\right]\right\}\\
    \times&\exp\left\{+\int_{2\pi/L}^{\infty} dq e^{-a q}\frac{\left(q^2+\varepsilon_{q}^\mathrm{c2}\right)m^2}{2q^2\varepsilon_{q}^\mathrm{c3}}
    \left(1-e^{-i\varepsilon_{q}^\mathrm{c}t}\right)\right\},
\end{aligned}\\
\begin{aligned}
    ig_\mathrm{c}^<(x,t)=\frac{\delta(x)}{2\pi a}&\exp\left\{-\int_{2\pi/L}^{\infty} dq e^{-a q} \frac{m^{4}}{4 q^{2} \varepsilon_{q}^\mathrm{c}}\left[\frac{1-e^{i \varepsilon_{q}^\mathrm{c}t}}{\left(\varepsilon_{q}^\mathrm{c}+q\right)^{2}}+\frac{1-e^{i \varepsilon_{q}^\mathrm{c}t}}{\left(\varepsilon_{q}^\mathrm{c}-q\right)^{2}}\right]\right\}\\
    \times&\exp\left\{+\int_{2\pi/L}^{\infty} dq e^{-a q}\frac{\left(q^2+\varepsilon_{q}^\mathrm{c2}\right)m^2}{2q^2\varepsilon_{q}^\mathrm{c3}}
    \left(1-e^{i\varepsilon_{q}^\mathrm{c}t}\right)\right\}.
\end{aligned}
\end{align}
After the Fourier transformation, we obtain the gauge violation spectral functions as show in the main text. Here $L=1000$ in Schwinger model, $L\rightarrow\infty$ in deconfinement model and $\alpha=0.01$ for all data have converged the results as illustrated in Fig.\ref{converge}. In the case $m=0$ or $g=0$, the Schwinger-like models reduce to massless Dirac model. All results above are consistent with the known results in the particular case.

\begin{figure}
  \centering
  \includegraphics[width=0.8\textwidth]{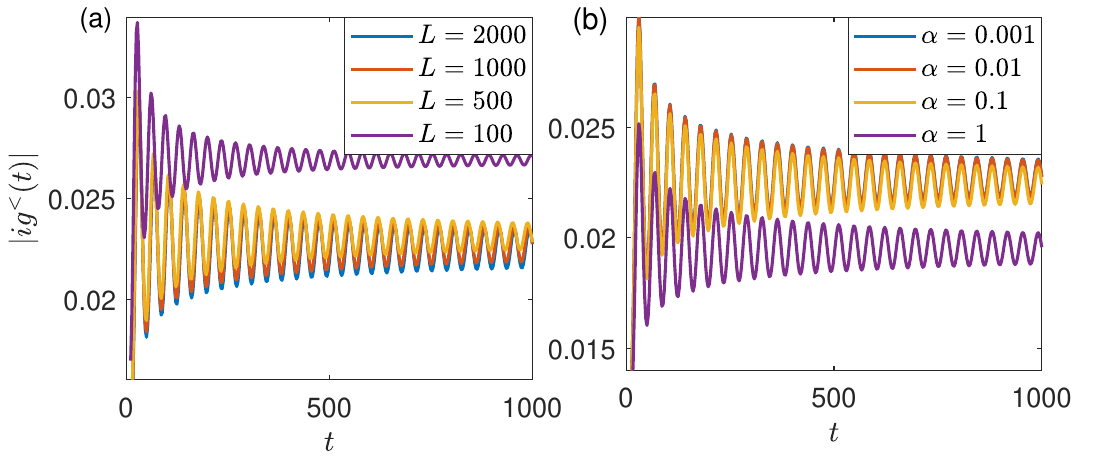}
  \caption{The norm of the gauge violation Green's function of Schwinger model with $m=0.17$. (a) Results for $L=100,500,1000,2000$ and $\alpha=0.01$. (b) Results for $\alpha=1,0.1,0.01,0.001$ and $L=1000$. We find $L=1000$ and $\alpha=0.01$ have converged the results.}\label{converge}
\end{figure}
\end{widetext}

\end{document}